\documentclass[twocolumn]{aastex631}
\usepackage[normalem]{ulem}

\shorttitle{An over-caffeinated search for coffee in space}
\shortauthors{Eistrup, Tychoniec et al.}
\graphicspath{{./}{figures/}}

\begin{document}

\title{Taurine in Taurus -- An Over-Caffeinated Search for Coffee in Space}

\author{Christian Eistrup}
\affiliation{IScreamCoffee, Leiden, Netherlands}

\author{Łukasz A. Tychoniec}
\affiliation{Brisman Kawowy Bar, Poznań, Poland}

\author{Iris Nijman}
\affiliation{Borgman \& Borgman, Leiden, Netherlands}

\author{Marta Paula Tychoniec}
\affiliation{Brisman Kawowy Bar, Poznań, Poland}

\author{Siroon Bekkering}
\affiliation{Dagger Coffee, Utrecht, Netherlands}

\author{Anna Gaca}
\affiliation{Brisman Kawowy Bar, Poznań, Poland}

\begin{abstract}

Caffeination can open tired eyes and enhance focus. Over-caffeination, furthermore, can lead to \textcolor{red}{\sout{errors}}, but also to unexpected !\textsc{discoveries}! that might not have happened without 30 hours of sleep deprivation and 500mg of caffeine in our bodies. This paper presents exactly such a discovery. Upon much staring into our coffee cups, empty anew, the thought struck us: coffee in space. Caffeine may not be the only key. HL Tau, Taurus, bull... Taurine! We grinded some red bourbon for a new pour-over, and developed the new, coffee-groundsbreaking Large Astrocomical Taurine Tester Experiment (LATTE) in just 1/4 of a day. We felt bull-ish about our chances of making a great discovery! We installed LATTE, aimed it at the well-known young star HL Tau, and there it was: an abundance of taurine gas beautifully outlining a cup of cosmic flat white, with the ring structure of HL Tau turning out to be latte art performed by a skillful cosmic barista. The first Robusta discovery of coffee in space. Speaking of coffee, we hope you have a nice hot cup with you, and we encourage you to pun-tinue all the way to the end of this bean-grinding paper.

\end{abstract}


\section{Introduction}

Astronomy is the science of the night sky. The field has always relied on the curiosity and perseverance of night hawks, who, in solitary and sometimes cold conditions have spent the early hours of the day in buildings with domes and telescopes inside, searching for long-black holes. Astronomical research needs the dark, but as humans are built for the day, humans have embraced the effects of focus-enhancing nutritional additives, first and foremost caffeinated drinks like coffee and tea.

As countless warm cups have bean consumed at astronomical observatories, the field of astrochemistry has led to the discovery of molecules in space that are closely linked with life as well know it. These molecules include ones essential for life on Earth \citep[water, H$_{2}$O, see][]{cheung1969water,ewinewater}, as well as ones that are connected with modern human life, such as the sugar glycolaldehyde \citep[CHOCH$_2$OH, see][]{jorgensen2012} and the alcohol ethanol \citep[CH${_3}$CH${_2}$OH, see][]{zuckerman1975}.

As the astrochemical search continues, astrochemists keep looking for spectral signatures of hitherto undiscovered molecular compounds in space \citep[see yearly updated list of known interstellar molecules,][]{mcguire2022living}, as well as a broader understanding of the reactive connections between these compounds: what are they made of, where are they made, and what complex chemistry could they possibly be processed into themselves.

Ironically, a molecule considered intimately linked to the overall success of astronomical research, namely caffeine (C$_{8}$H$_{10}$N$_{4}$O$_{2}$), has yet to be discovered. This is not to say that caffeine does not exist in space, but rather that potential coffee beans harboured in planet-forming disks have not yet drifted close enough to their host stars to roast, nor have they undergone the well-known process of collisional fragmentation \citep[see, e.g.,][]{krijt2015} which would serve to powderise, and potentially even vaporise the caffeine carried in the beans, thereby making the caffeine spectrally detectable. 

Given that coffee beans on the Earth only ripen once a year, and for a relatively short period of time (2-3 months out of 12), a seasonal effect may be expected in space as well: the duration of the local year on a given exoplanet in formation in a protoplanetary disk may help determine when the exoplanetary coffee beans are ripe. In combination with theoretical insights into the radial drift speeds \citep[see][]{brauer2008,lambrechts2012} of coffee beans, their caffeinated collisional behaviour, and the future travel of caffeine-discovering goats to space, it may be possible to deduce optimal periods of time during an Earth year in which a combination of ALMA, VLA and GBT efforts could lead to a molecular discovery. Future space travelers who bring along goats for coffee discovery, but who encounter space tea instead of coffee, may even develop thoughts about growing goat-teas as means of facial decoration.

One more conventional option is to look for traces of nitro coldbrew. However, its possible existence around young stars is observationally limited, as N$_2$ gas has no permanent dipole moment. Observing nitro coldbrew therefore depends on the local N$_2$ isotope fractionation, which renders this type of cold brew non-optimal for initial coffee discovery.

This paper presents an alternative, focus-enhanced approach to finding coffee in space. While caffeine has built a gigantic following on the Earth, other molecular compounds that are naturally occurring in human bodies feature similar, if not stronger necessity-of-sleep-reducing effects. Case in point: the amino acid taurine (C$_2$H$_7$NO$_3$S). For astronomers, and human beings generally who were introduced to coffee through the kind of coffee\footnote{For clarity, a more precise description of average-quality ``coffee'' at astronomical institutes is ``caffeinated, dark, lukewarm water’’} offerings available at a typical astronomical institute, this coffee experience is likely to have led to aversion towards coffee. This aversion is somewhat similar to the reaction that non-Dutch and non-Danish adults have to tasting salty licorice for the first time.

For astronomers not drinking coffee, who want a stronger focus-enhancing effect than what is usually achieved drinking typical teas (Yerba Mate and similar types excluded), many go for so-called energy drinks. These drinks contain different compounds that are associated with enhanced performance, such as sugars, caffeine, and, importantly, taurine. 

Taurine, one of the semi-essential amino-acids, is named after the ox (taurus in latin), because it was originally isolated from the biles of oxen. It is unique to mammals, and particularly immune cells are high in taurine \citep[][]{taurine}. Un-surprisingly, taurine is not only responsible for the ability to sit and have your eyes wide open. The deficiency of taurine causes renal dysfunction, hyperinflammation and loss of retinal photoreceptors \citep{ripps.shen.2012}, subsequently leading to the belief that all the stars, including the Sun, have exploded, someone cut the electricity at the lab and it's so dark I can't find my coffee. (Ah, I smell it now. Phew!) \\

\emph{This insight is the clue that inspired this paper!}
\\\\
Instead of focusing on observing the directly coffee-related caffeine in space, this paper pursues the discovery of taurine. As indeed the name indicates, taurine is linked to the ox, taurus, which, coincidentally, is a famous constellation in the night sky. Furthermore, the constellation Taurus is home to the young star HL Tau, which hosts a well-studied protoplanetary disk \citep{hltau}. This made the HL Tau system immediately desirable to study for spectral signatures of taurine.

\section{Method}

With the state-of-the-art Large Astronomical Taurine Tester Experiment (LATTE), we have obtained an observation of the HL Tau disk. This pioneering instrument involves brewing a coffee with water which was subsequently exposed to the incident light from the observational target.

The quality of the observation is strictly proportional to the quality of the coffee served to the artist and the reproducibility of the observation is a cornerstone of the scientific method. Therefore, below we provide a detailed description of the extraction process.

\subsection{Materials}
We chose El Salvador arabica coffee, Red Bourbon varietal, coming from the Rudolfo Battle coffee farm placed in the Santa Ana region. The coffee underwent a natural processing at the farm. It was roasted on 7th of March 2022 by the Manhattan Coffee Roasters in The Netherlands. 

\subsection{Coffee extraction}
We grinded 31 grams of coffee with the Comandante Grinder featuring the enhanced red clix tool (Comandante, Germany) on 45 clix. Water, filtered with a home-grade carbon filter showed an ion fraction of 384 ppm\footnote{A clear signature of the fact that the filter should be replaced - hope someone port-a-filter!}, and was heated up to 93$\rm{^o}$C. In accordance with \cite{angeloni.guerrini.2019}, the V60 filter method was selected because of a decent ratio of bioactive compounds-to-water, and time spent on the LATTE-instrument. In the experiment, we used V60-02 paper filters (Hario, Japan). After initial pre-infusion with 45 mL of water, a total of 500 mL of water was poured over in a total time of 204 s. Subsequently, the resulting perfect-blend beverage was served to the artist who provided the rendering of the targeted object (Fig.  \ref{fig:my_label}).

\section{Results}

\begin{figure}
    \centering  
    \includegraphics[width=0.47\textwidth]{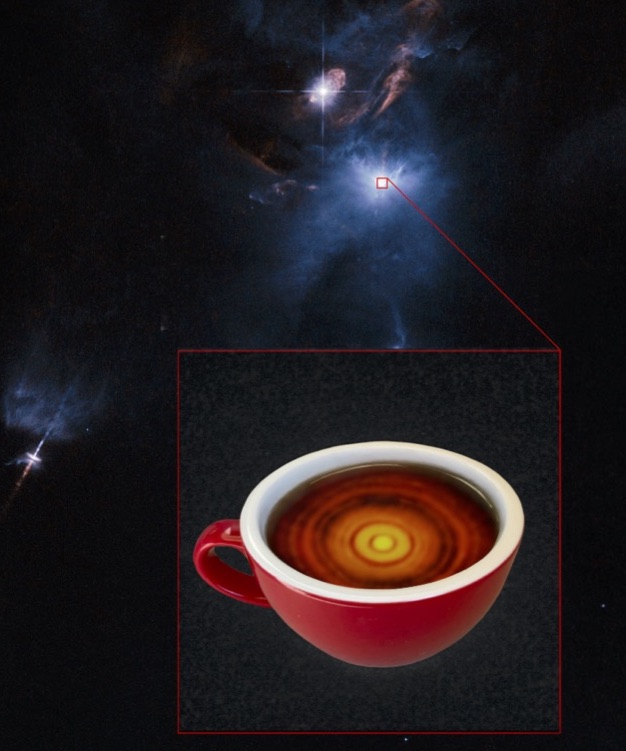}
    \caption{A zoom-in on the star-forming region in Taurus, which harbours the young star HL Tau, with its characteristic rings. Using the coffee-groundsbreaking LATTE-instrument, taurine has become detectable. The spatial structure of the taurine clearly outlines a red coffee cup, complete with white rim and handle. The ring structures of HL Tau, hitherto thought to be results of exoplanets in formation, are, in fact, latte art, as performed by a skillful cosmic barista. With this, LATTE presents the first evidence of coffee in space! Adapted based on images credited to: ALMA (ESO/NAOJ/NRAO), ESA/Hubble and NASA, Judy Schmidt.}
    \label{fig:my_label}
\end{figure}

Fig. \ref{fig:my_label} presents the results of the experiment. The ringed structure of the HL Tau disk, for years associated with various processes like planet formation \citep[e.g.,][]{Takahashi.Inutsuka2016}, appears to be a cosmic latte art. The key evidence is the cup revealed at the narrow frequency range of the LATTE-instrument. A red colour of the cup could be an indication of the cup moving away from the observer.

It should be noted that the latte art pattern observed is not commonly seen in the Earthly coffee houses. It is indeed difficult to create a pattern that would result in a ring-shaped structure with such a high contrast, using any free-pour method. We therefore hypothesize that barista etching has to have bean involved - truly a signature of a ``stellar'' cosmic barista at work!

\section{Discussion}

\subsection{Astronomical implications}
It needs to be further explored if other astronomical structures are results of a similar phenomenon, such as the accretion streamers observed in SU Aur system \citep{Ginski.Facchini.ea2021} or whether our Galaxy's name Milky Way indeed has some links to actual milk. If so, it will be interesting to understand if this milk is early-stage, conventional cow milk, late-stage, modern types such as oat, soy, or almond, or maybe a mix. A further hunt for lactose, which traces cow milk, starch, which traces oatmeal(k), and soy in the Galactic plane is therefore recommended. It is noted that astronomers searching for almond milk will face risks of going nuts, and so observers with allergies should take precautions. In addition, these searches could also lead to deja-brews, including the feeling from the 2nd Coffee Wave that a Pumpkin Spice Latte qualifies as coffee. Further important questions still remain to be explored, for example: how well would the cosmic flat white pair with the space pretzel \citep{Alves.Caselli.2019}?
 
 \begin{figure}
        \includegraphics[width=0.5\textwidth]{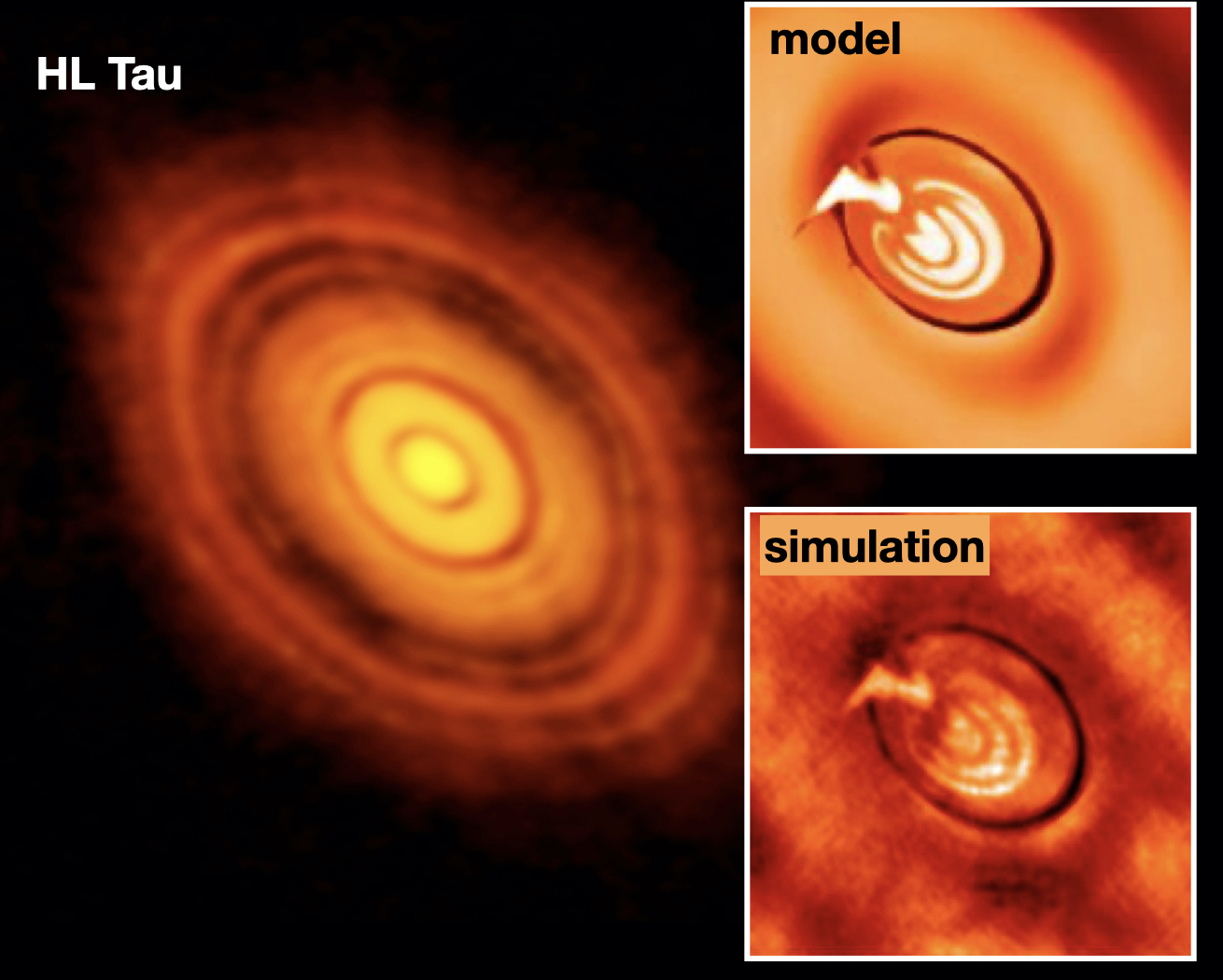}
    \caption{A simulated ALMA observation of latte-art pattern in the innermost ring of the HL Tau. HL Tau image credit: ALMA (ESO/NAOJ/NRAO).}
    \label{fig:label2}
\end{figure}
 The gas-phase spectral signatures of taurine has bean verified in the 6-14 GHz regime \citep{Cortijo.Sanz.ea2009}. The VLA observations covering this frequency range \citep{Wilner.Ho.ea1996} do not have enough spatial resolution to reveal if the latte art pattern is recovered at those frequencies as well. However, we investigate the possibility of discovering another latte-art pattern in the innermost disk of HL Tau.
 
 We simulated a model latte-art pattern commonly known as a tulip into the innermost ring of the HL Tau disk. Due to the fact that this is a young protostellar object we expect the latte art to still be ongoing formation, hence we also model the pouring milk filament. We use \verb simalma task in CASA software \citep{McMullin.Waters.ea2007}. The ALMA Cycle 9 configuration C-9 in Band 8 (460 GHz) were simulated to target HL Tau disk with an additional pattern for 7200 s which resulted in an observation presented in Fig. \ref{fig:label2}. High contrast and symmetry of the pattern clearly poses ALMA as a contender in the next Latte Art Championships.

\subsection{Wider societal impacts}

The discovery of taurine as presented in this paper is not only one more amino acid discovered in space, but also reassuring in the context of human space travel, especially considering its strong anti-inflammatory properties, which will be much needed when humans come in contact with unknown space bacteria. It is imperative to the survival of humans in space that we can gather resources and nutrition \emph{en-route} increasing our chances of staying awake and focused, and surviving.

In addition, while humankind is fortunately still able to be surprised with, and indulge in the beauty of Earthly nature, it is indeed spectacular how an outcome of the laws of nature is producing features reminiscent of what humans call ``A Flat White with Latte Art’’. Hopefully, the beauty of the Universe that humans continue to uncover will serve as a reminder for humankind that we have a wonderful planet to live on, and that we need to care for it, if it is to last.

Lastly, with this paper, the authors hope to inspire astronomical and other research institutions to take their coffee and hot beverage situations seriously. A good coffee or a nice cup of tea can and should serve as both caffeination/refreshment, but more importantly as a nice element of self-indulgence and self-care, especially in a work-setting. Going for a coffee break with colleagues is much more delightful when the main topic of conversation during the break \emph{is not} how bad the coffee tastes. The happiness and positivity that a nice cup of coffee induces is quintessential for a productive and vibrant working environment.

\section{Conclusion}
We are over the moon to have discovered the first evidence of coffee in space. Thanks to the coffee-groundsbreaking LATTE instrument, we show that the famous HL Tau system is, in fact, a cosmic flat white with latte art and barista etching. We consider this an im-pressive (at $\sim9$ bars\footnote{Appropriate for espresso extraction}) achievement, which would had not bean possible without our inter-disciplinary approach, joining insights from food science (AG), astrochemistry (CE \& ŁT) science communication (IN), immunology (SB), and the arts (MPT).

As coffee in space is exciting, the enhanced focus we all gain from this discovery can be used constructively to the benefit of astronomers (and all scientists alike) at all institutions. With this, we kindly apply some French press-ure to suggest that research institutions review and improve their coffee offerings.

While an in-house barista and in-house coffee roasting would be appreciated, a number of smaller steps can be taken to im-brew the coffee offered: 

\begin{enumerate}
    \item Freshness of the coffee beans, generally defined as freshly roasted, ground within a few weeks after roasting, immediately extracted and consumed, is proven to have a dramatic effect on the quality of the beverage \citep{Yeretzian.Blank.ea2017}.
    \item Using lighter rather than darker roasts in order to ensure pronouncing of the natural flavours of the beans rather than burned aroma and having higher caffeine content, thus higher anti-inflammatory properties \citep{Hecimovic.BelscakCvitanovic.ea2011}.
    \item Seeking out smaller-scale and fair trade coffee suppliers.
    \item Keeping the coffee machine clean.
    \item Changes in temperature after brewing also influences on coffee brew flavour. Keeping drinking coffee in a slow pace may surprise due to aroma changing \citep{adhikari.chambers.2019}.
    \item There is no perfect brewing method. But there is always a space to compare with the others.
    \item Trying to experiment. Even basket geometry can change a coffee flavour \citep{frost.ristenpart.2019}. 
    \item Thinking about horrible coffee (e.g. at the conference) in a good way may improve their overall acceptance \citep{bemfeito.guimaraes.2021}. 
\end{enumerate}

The metropolitan coffee culture all the way back in the 16th century in Arabia can serve as inspiration, by reminding us that coffee houses were once places for scholars and intellectuals to meet to discuss the topics of those days. Back then, coffee houses were even known as ``Schools of the Wise''\footnote{https://www.ncausa.org/about-coffee/history-of-coffee}. 

We therefore kindly suggest that academic meeting organisers consider booking a local coffee house for their next meeting, and let the participants indulge in the magic of the barista!

{\it Acknowledgements.} The authors thank the research institutes that have already improved their coffee offerings. ŁT acknowledges Brisman Kawowy Bar in Poznań for the support in pursuing his barista career and Chummy Coffee in Leiden for being a crucial moral support during the process of writing a PhD. SB acknowledges Blommers Coffee Brewers - location Radboudumc for brewing her specialty coffee at work and Blackbird Bio Kaffee in Utrecht for introducing her to quality coffee 10 years ago. \\
We further highlight that there is war in Europe, and that we are privileged to be able to work, continue our daily lives, and write about coffee, while Ukrainians are fleeing their country and people, including children, are suffering and dying. If you have made it to these acknowledgements, please consider a donation to help the people of Ukraine: \href{www.comebackalive.in.ua}{www.comebackalive.in.ua}
\\

\bibliography{main}{}
\bibliographystyle{aasjournal}

\end{document}